\documentclass[prb,showpacs,preprintnumbers,preprint,superscriptaddress]{revtex4}

\usepackage{epsfig}
\usepackage{graphics}
\usepackage{subfigure}
\usepackage{amsmath}
\makeatletter
\@ifundefined{textcolor}{}
{%
 \definecolor{BLACK}{gray}{0}
 \definecolor{WHITE}{gray}{1}
 \definecolor{RED}{rgb}{1,0,0}
 \definecolor{GREEN}{rgb}{0,1,0}
 \definecolor{BLUE}{rgb}{0,0,1}
 \definecolor{CYAN}{cmyk}{1,0,0,0}
 \definecolor{MAGENTA}{cmyk}{0,1,0,0}
 \definecolor{YELLOW}{cmyk}{0,0,1,0}
}

\usepackage{epsfig}\usepackage{graphics}\usepackage{subfigure}
\def\e{\begin{equation}}
\def\f{\end{equation}}
\def\_#1{{\bf #1}}

\def\E{\varepsilon}
\def\M{\mu}

\def\.{\cdot}

\makeatother

\begin{document}

\title{Towards negative index self-assembled metamaterials}

\author{M. Fruhnert\footnote{martin.fruhnert@uni-jena.de}}
\affiliation{Institute of Condensed Matter Theory and Solid State Optics, Abbe Center of Photonics, Friedrich-Schiller-Universit\"at Jena, Max-Wien-Platz 1, 07743 Jena, Germany}

\author{S. M\"uhlig}
\affiliation{Institute of Condensed Matter Theory and Solid State Optics, Abbe Center of Photonics, Friedrich-Schiller-Universit\"at Jena, Max-Wien-Platz 1, 07743 Jena, Germany}

\author{F. Lederer}
\affiliation{Institute of Condensed Matter Theory and Solid State Optics, Abbe Center of Photonics, Friedrich-Schiller-Universit\"at Jena, Max-Wien-Platz 1, 07743 Jena, Germany}

\author{C. Rockstuhl}
\affiliation{Institute of Condensed Matter Theory and Solid State Optics, Abbe Center of Photonics, Friedrich-Schiller-Universit\"at Jena, Max-Wien-Platz 1, 07743 Jena, Germany}
\affiliation{Institute of Theoretical Solid State Physics, Karlsruhe Institute of Technology, Wolfgang-Gaede-Strasse 1, 76131 Karlsruhe, Germany}


\begin{abstract}
We investigate the magnetic response of meta-atoms that can be fabricated by a bottom-up technique. Usually such meta-atoms consist of a dielectric core surrounded by a large number of solid metallic nanoparticles. In contrast to those meta-atoms considered thus far, we study here for the first time hollow metallic nanoparticles (shells). In doing so we solve one of the most pertinent problems of current self-assembled metamaterials, namely implementing meta-atoms with sufficiently large resonance strength and small absorption. Both conditions have to be met for deep sub-wavelength meta-atoms to obtain effectively homogeneous metamaterials which may be meaningfully described by negative material parameters. Eventually we show that by using these findings self-assembled negative index materials come in reach.
\end{abstract}

\maketitle

\section{Introduction}

A requirement for many applications in the fields of metamaterials (MMs) and transformation optics is the availability of bulk materials that possess a strong response to the magnetic field in the visible and infrared (IR) spectral domain \cite{Capolino_CRC,Yannopapas_JP,Scharf_PRL,Simovski_PRB}. Whereas quasi-two-dimensional structures \cite{Liu_Nat,Reinhardt}, i.e. meta-surfaces, are nowadays well-established and sustain such a magnetic response, it still remains a challenge to fabricate such materials as bulk structures with available top-down technologies. Consequently, bottom-up approaches have been suggested \cite{Capasso_OE,Scharf_AM,Barois_L,Macdonald_AM} that directly provide bulk materials. However, the advancement of this field strongly depends on the identification of meta-atoms that provide a suitable optical response \textit{and} which are amenable for a bottom-up fabrication.

Recently, it has been demonstrated that a highly isotropic magnetic response can be observed while relying on core-shell clusters as the meta-atom. These clusters consist of a dielectric core sphere covered by a huge number of plasmonic nanospheres forming an effective shell. At a particular frequency, an effective current can be excited in the shell flowing around the core sphere. This causes a scattered field identical to that of a magnetic dipole. There are multiple concepts of utilizing these core-shell structures as meta-metamaterials \cite{Dintinger_OME,Yannopapas_prb,Scharf_PRL}, which consist of periodically arranged blocks build of meta-atoms, homogeneous core-shell spheres \cite{Paniagua_NJP,Paniagua_SR,Morits_JO} and core-shell clusters \cite{Alu_OE,Capolino_OE,Albani_OE,Yannopapas_rrl,Dionne_nl}. The main advantage of these core-shell clusters is their isotropic response when compared to ordinary meta-atoms such as split-rings \cite{Liu_Nat} or cut-plate pairs \cite{Wegener_OL}. Furthermore, these core-shell clusters can be fabricated by self-assembly techniques which allow to produce bulk materials at short time, large amounts, and low costs \cite{Muehlig_ACSN}.

The important step from an isolated meta-atom with a magnetic dipole response (e.g. a single core-shell cluster) to a true MM, consisting of many and possibly densely packed meta-atoms, consists in assigning meaningful effective material parameters, e.g.\ an effective permittivity and permeability. Their unambiguous assignment, however, requires a local response to an external electromagnetic field and thus deep sub-wavelength meta-atoms. For cut-plate pairs it has been only recently shown that this can be achieved by the exploitation of an extreme coupling regime. As a result of this strong coupling the magnetic dipole resonance of cut-plate pairs could be shifted to the near IR while maintaining their small geometrical extensions \cite{Menzel_OL}.

Moreover, a referential benchmark concerning material properties is the availability of a negative permeability. Although not being strictly necessary to obtain a negative index material, it is usually highly beneficial \cite{veselago}. A negative permeability is only in reach for a sufficiently strong magnetic resonance. Moreover, having such a resonance available the material may be even operated off-resonantly. This entails a reduced absorption whereas the dispersion remains sufficiently strong, such that all anticipated effects for a negative permeability appear much stronger and are less affected by absorption. Thus, the stronger the resonance the better. One solution to this problem is the incorporation of gain material into the unit cells \cite{Campione_ome}. However, since the experimental problems remain, solutions are highly desirable that only require a modification of the geometry of the meta-atom.

Here, we suggest potential solutions to these challenges. Specifically, we show how it is possible to shift the magnetic dipole resonance of core-shell clusters into the near IR while maintaining their spatial dimensions. To this end we exploit hollow metallic nanospheres as constituents of a novel meta-atom. This is in contrast to ordinary meta-atoms where primarily solid metallic nanospheres have been considered thus far. Our design reveals a magnetic response of a core-shell cluster with deep sub-wavelength dimensions. Moreover, thanks to the smaller radiation losses stipulated by the smaller size as well as the lower intrinsic absorption at near IR frequencies, this approach results in ultra-strong magnetic resonances and eventually in an effectively homogeneous and isotropic material with a negative permeability. By modifying the meta-atoms such that they equally sustain a sufficiently strong electric resonance at the same frequency, a negative index material can be eventually achieved.

With respect to a practical realization of these meta-atoms we also study the effect of disorder of the plasmonic nanospheres forming the shell. It is shown that our design is sufficiently robust against disorder, rendering its fabrication feasible with bottom-up and self-assembly methods.

\section{Methodology}

\begin{figure}[htbp]
\centering \includegraphics[width=0.8\textwidth]{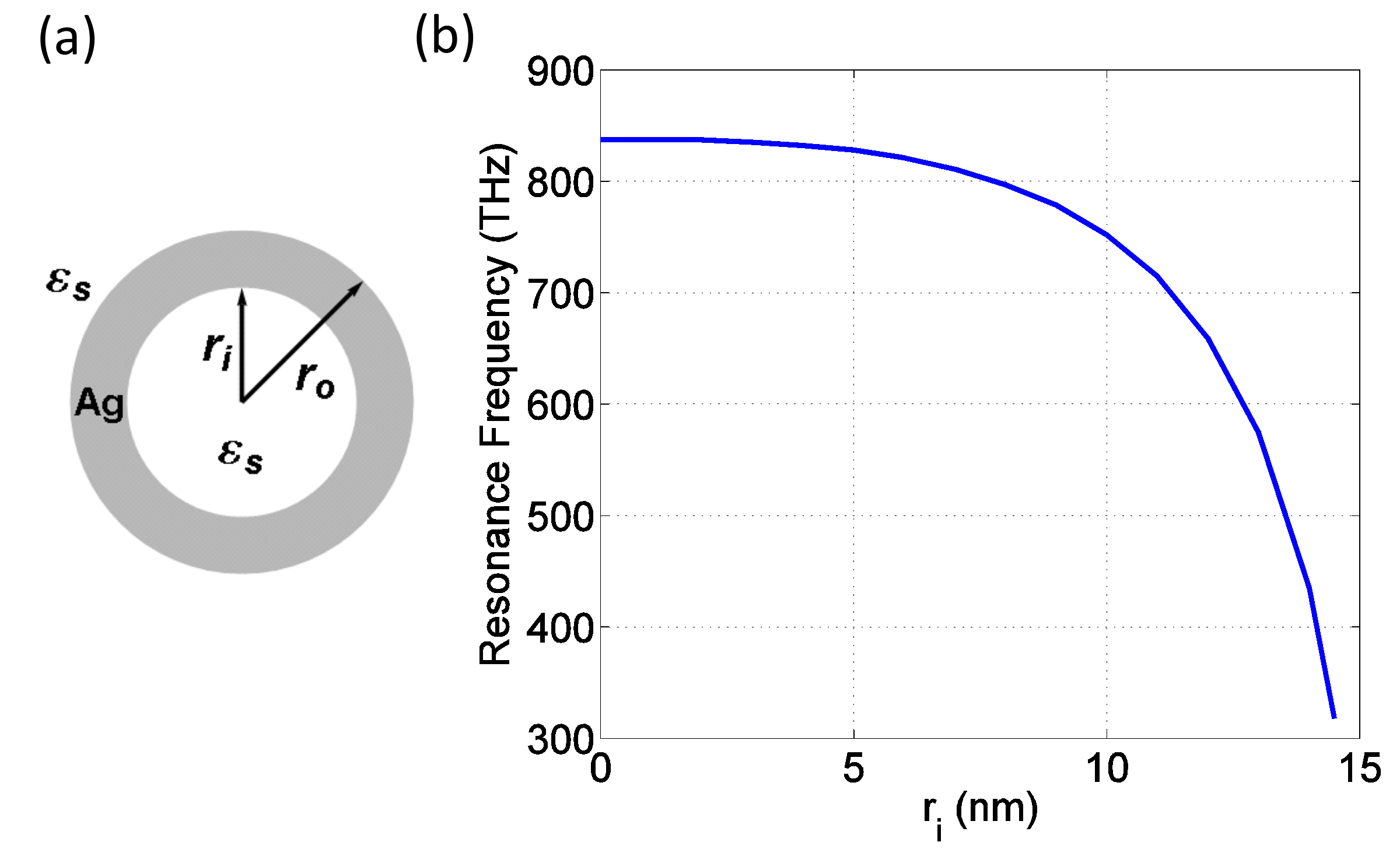} \caption{Silver shell nanoparticle. a) Geometry. b) Resonance frequency as a function of the inner shell radius $r_{\mathrm{i}}$ with a fixed outer radius $r_{\mathrm{o}}=15$ nm.}
\label{fig_shell}
\end{figure}

All numerical results presented in this work rely on a self-consistent solution to the scattering problem of light at an arbitrary number of spherical particles for a given external illumination. The spheres are allowed to be made form different materials and may have different size. Ultimately, not just bulk spheres are considered but more complicated radial geometries, i.e.\ spheres consisting of multiple shells of different materials, can easily be accommodated. A multi-scattering algorithm based on \cite{Xu} has been used for this purpose.

At its core, it solves the Mie scattering problem for an isolated sphere \cite{Muehlig_PRB}. To this end the total electromagnetic field outside the sphere is decomposed into an incident and a scattered field whereas the field inside the sphere is just the internal field. All these fields are expanded into vector spherical harmonics $\textbf{N}_{nm}(r,\theta,\phi)$ and $\textbf{M}_{nm}(r,\theta,\phi)$. For example the expansion for the scattered electric field reads as:

\begin{align}
\textbf{E}_{\mathrm{sca}} & (r,\theta,\phi,\omega)=\sum_{n=1}^{\infty}\sum_{m=-n}^{n}k^{2}E_{nm}\cdot\\
 & \left[a_{nm}(\omega)\textbf{N}_{nm}(r,\theta,\phi,\omega)+b_{nm}(\omega)\textbf{M}_{nm}(r,\theta,\phi,\omega)\right].\nonumber
\end{align}
Here, $k^{2}=\left(\omega^{2}/c^{2}\right)\E(\omega)\M(\omega)$ is the dispersion relation in the surrounding medium with the propagation constant $k$, the permittivity $\E(\omega)$ and the permeability $\M(\omega)$ .

The exact contribution of all these vector spherical harmonics to the respective field, expressed by the expansion coefficients $a_{nm}$ and $b_{nm}$, can be calculated by enforcing the boundary conditions to the angular components of the electric and magnetic field at the spherical surface. The expansion coefficients are known as the scattering coefficients. Closed form analytical expressions exist for single spheres and an extension towards spherical objects with radially varying material parameters can be easily performed.

Provided that the center of the coordinate system coincides with that of the core object the associated coefficients have an unambiguous physical meaning. The $a_{nm}$ coefficients express how electric multipole moments contribute to the scattered field, whereas the $b_{nm}$ coefficients determine the contribution of the magnetic multipole moments. For an even more intuitive physical interpretation, they can be easily transformed into Cartesian multipole moments \cite{Muehlig_meta}. Then, it can be seen that the $a_{1m}$ and $b_{1m}$ components with $m=-1,0,1$ emerge in the electric and magnetic dipole moments, respectively.
From these dipole moments, the electric and magnetic polarizability $\alpha_{\mathrm{E}}$ and $\alpha_{\mathrm{M}}$ can be easily obtained by normalizing the dipole moments with the amplitude of the incident field. A scalar treatment for the polarizability is fully sufficient since the optical response will be isotropic. This polarizability can be used to calculate effective material parameters, as e.g. the effective permeability by using the Clausius-Mossoti equation
\e
    \M_{\mathrm{eff}}(\omega)=\M_\mathrm{s} \frac{3+\frac{2 N \alpha_\mathrm{M}(\omega)}{V}}{3-\frac{N \alpha_\mathrm{M}(\omega)}{V}},
    \label{eq_CM}
\f

where $\M_{\mathrm{s}}=1$ is the permeability of the environment, $N$ is the filling fraction and $V$ the volume of the unit cell.
Technically, our calculations have been performed with a fourth-order multipole expansion. Thus, in the numerical solution we are considering multipole contributions up to a 16-pole, though the dominating contributions to the scattering cross section are the dipole and quadrupole coefficients as we will see in Fig.\ \ref{fig_multipoles}.
The basic building block of the present meta-atoms consist of spherical particles with a metallic shell and a hollow dielectric core. This is in contrast to the usually considered bulk spheres. We used experimentally determined values for the material parameters of silver \cite{Johnson_Christy} and $\E_\mathrm{s}=1.7$ for the dielectric core of the nanoshells. A bulk sphere made of silver possesses a resonance at about $850$ THz. There, a localized surface plasmon polariton (LSPP) is excited where the density oscillation of the free charges is resonantly coupled to the external electromagnetic field. This resonance at a relatively high frequency constitutes a problem to obtain sub-wavelength resonances for the final meta-atom. Therefore, this resonance has to be shifted to longer wavelengths to achieve a truly sub-wavelength structure. This is possible by using silver nanoshells. These shells exhibit a red-shifted resonance that is stronger the thinner the nanoshell will be \cite{Yannopapas_oqe,Halas_S,Halas_NL}. The red-shift occurs due to a hybridization of the LSPP that is supported at the interface between the metallic shell and the dielectric surrounding and the interface between the metallic shell and the dielectric core. In the following we consider nanoshells with inner and outer radii $r_{\mathrm{i}}$ and $r_{\mathrm{o}}$, respectively, as shown in Fig.\ \ref{fig_shell}a), in a dielectric environment with $\E_\mathrm{s}=1.7$.

The LSPP resonance frequency of a single nanoshell as a function of the inner shell radius $r_{\mathrm{i}}$ for a fixed outer radius $r_{\mathrm{o}}$ is shown in Fig.\ \ref{fig_shell}b). It can be clearly seen that the frequency down shift amounts to about $500$ THz while increasing the inner shell radius from $0$ to $14$ nm, thus decreasing the shell thickness. This is possible while retaining the outer dimensions of the particle. Even though it sounds challenging to fabricate such thin silver shells, they are nowadays available with nanochemical methods that are based on the deposition of silver ions on electrostatically charged particles \cite{liu} or galvanic replacement methods \cite{prevo}.

In the following we exploit such tuneability to design deep sub-wavelength meta-atoms with a huge magnetic dipole response.

\section{Core-Shell Clusters}

\begin{figure}[htbp]
\centering \includegraphics[width=0.8\textwidth]{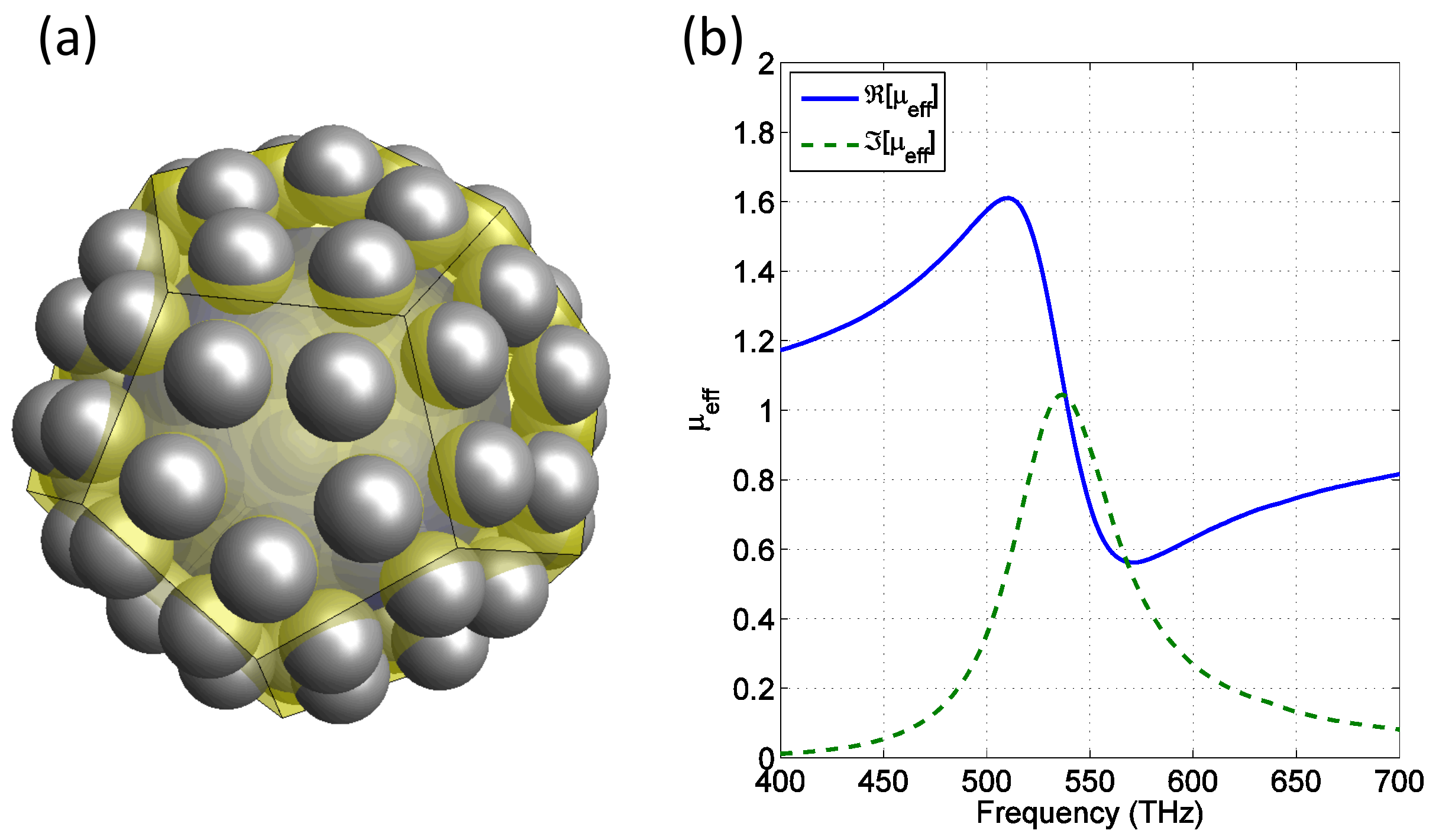} \caption{a) Sketch of the core-shell cluster where nanoparticles are arranged on the surface of a dodecahedron. b) Real and imaginary part of the effective permeability of a core-shell structure made from solid silver nanospheres.}

\label{fig_coreshell}
\end{figure}

The referential design is a core-shell cluster with $60$ solid nanospheres because, based on previous experiences of self-assembled metallic spheres on dielectric cores, the number of particles has to be sufficiently large. Moreover, motivated by specific fabrication techniques that exploit electrostatic interaction between the metallic nanospheres (which is repulsive since the nanospheres can be made negatively charged) \cite{dong_chem}, a reasonable assumption for the spatial arrangement of the nanospheres on the surface of the dielectric core is to arrange them in groups of pentamers at the facets of a dodecahedron, as sketched in Fig.\ \ref{fig_coreshell}a). In passing we note that this is just a referential nominal geometry and, of course, the eventual structure that is fabricated is not necessarily exactly characterized by such a geometry; although it resembles it closely. To take fully into account deviations from such nominal geometry, we consider later the effect of disorder on the design. The radius of the dielectric core sphere is $60$ nm and it has a permittivity of $\E_\mathrm{c}=2.25$. Initially we consider the nanospheres to be solid with a radius of $15$ nm. They consist of silver and the surrounding material is a dielectric with $\E_\mathrm{s}=1.7$.

From Fig.\ \ref{fig_coreshell}b) it can be seen that the magnetic dipole response of these clusters appears at $540$ THz. The effective permeability was calculated with the Clausius-Mossotti equation (\ref{eq_CM}), assuming a sufficiently large filling fraction of the bcc-lattice to ensure a significant effect. The permeability has a reasonable dispersion but the real part does not attain values less than $0.5$. The values compare to other designs that rely on solid nanospheres. In most cases this is insufficient for a negative index.

\begin{figure*}[htbp]
\centering \includegraphics[width=0.8\textwidth]{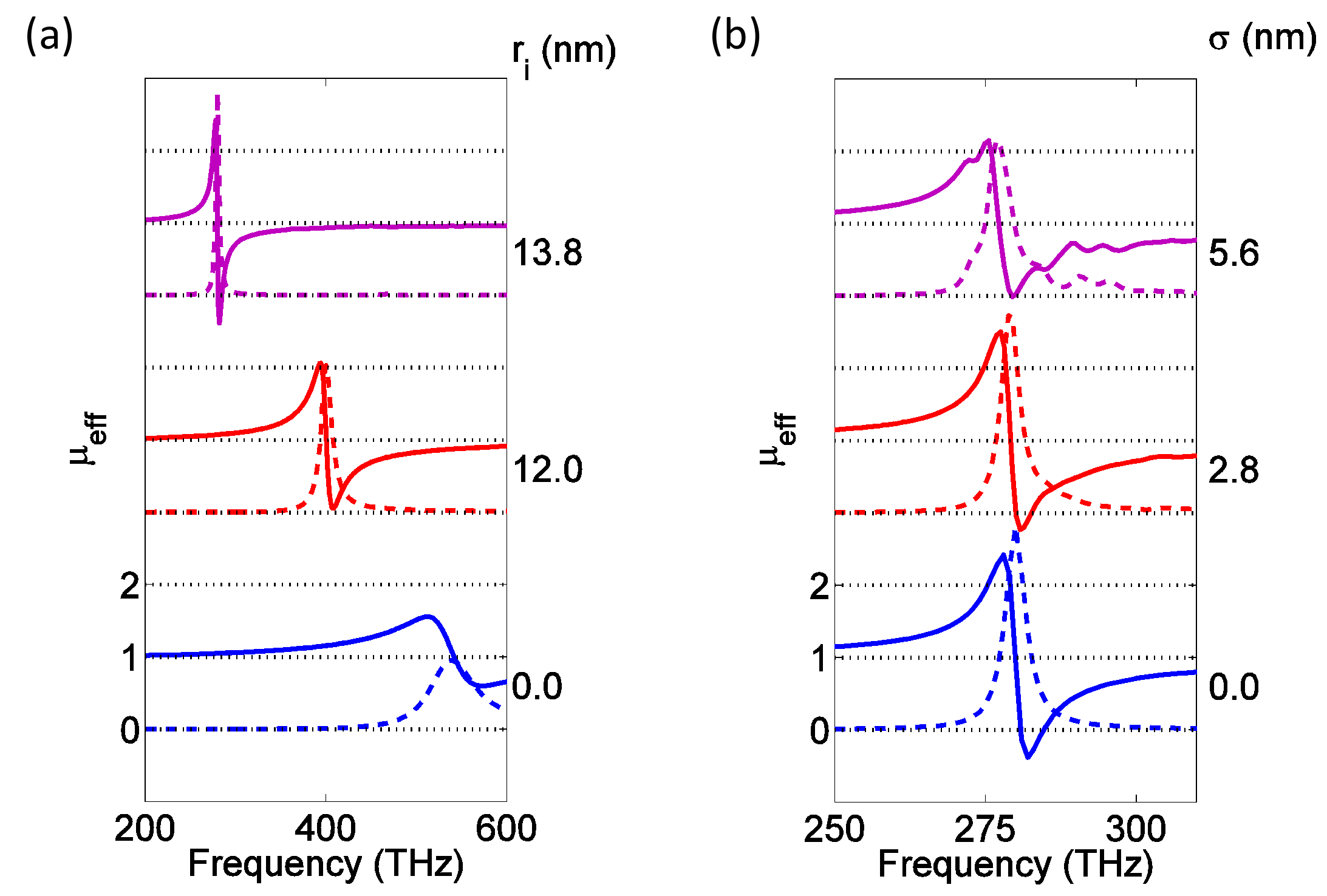} \caption{a) Effective permeability of the core-shell cluster made of nanoshells with different inner radii $r_{\mathrm{i}}$. The resonance shifts to smaller frequencies and gets sharper with increasing inner radius. b) Effective permeability of a cluster with $r_{\mathrm{i}}=13.8$ nm and normally distributed spatial disorder of the nanoshells, quantified by a average spatial displacement $\sigma$.}

\label{fig_waterfall}
\end{figure*}

Now the main idea consists in shifting the resonance of the polarizability to lower frequencies and to enhance its strength by replacing the solid nanospheres by silver nanoshells to form the core-shell cluster. As discussed above, the outer radius $r_{\mathrm{o}}$ of the nanoshells is kept constant while the inner one $r_{\mathrm{i}}$ is increased. This fully preserves the geometrical size of the core-shell cluster. The effect on the magnetic response for different inner radii is shown in Fig.\ \ref{fig_waterfall}a). Two effects can be recognized. First, and this is the expected one, the magnetic resonance shifts to lower frequencies into the IR. This is obvious, since the LSPP of the nanoshells dictates the resonance frequency of the effective current of the shell that causes the magnetic dipolar response. Second, the resonance features are sharpened upon this shift and additionally the permeability may eventually assume negative values. The stronger dispersion can be explained by the overall smaller relative size of the meta-atom, which lowers the radiative losses, and the reduced amount of metal in the nanospheres, which lowers the absorption. Note that this is only valid to some extend, because the loss can increase again in extremely thin shells due to the smaller mean free path of the electrons in the metal shell.

\begin{figure*}[htbp]
\centering \includegraphics[width=\textwidth]{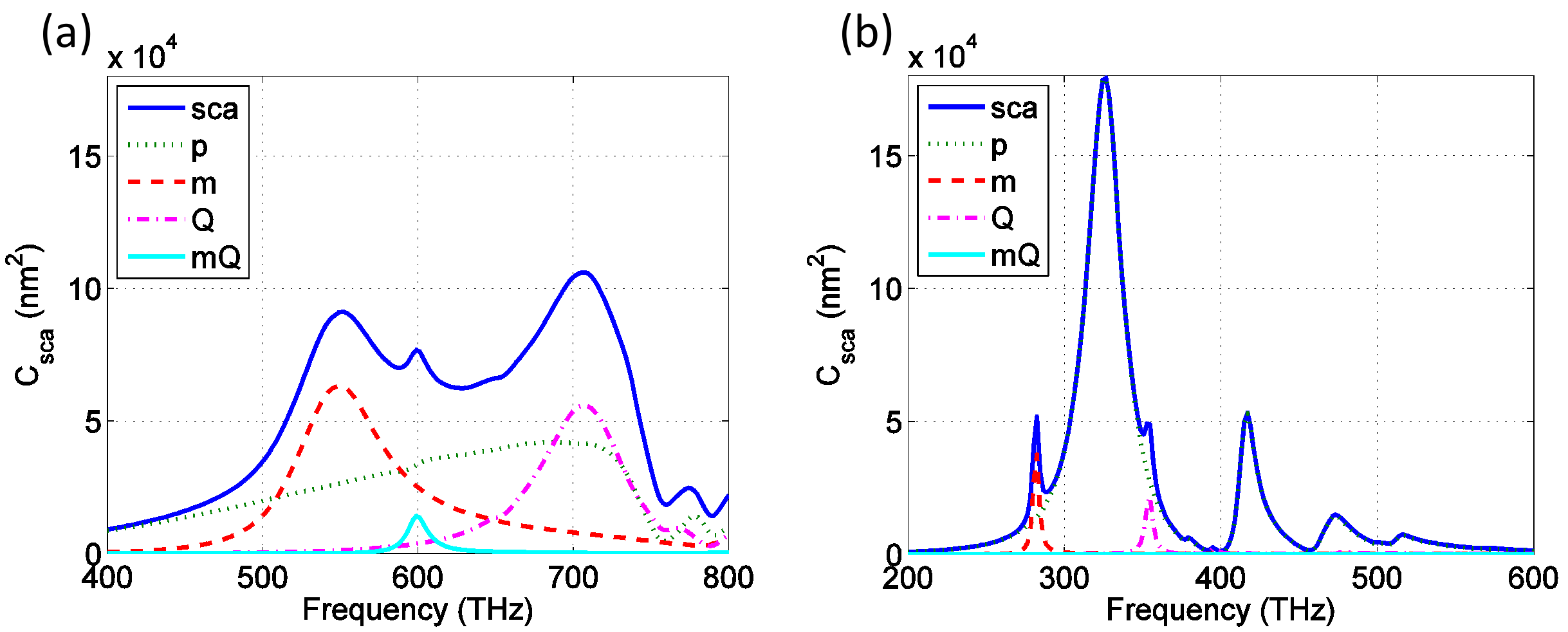} \caption{Scattering cross section of the core shell cluster. a)Bulk silver nanospheres. b) Silver nanoshells with $r_{\mathrm{i}}=13.8$ nm. Blue line: total scattering cross section. The other curves display the contributions of different multipoles: electric dipole (green dotted), magnetic dipole (red dashed), electric quadrupole (magenta dash-dotted), magnetic quadrupole (cyan)\protect \\
 }

\label{fig_multipoles}
\end{figure*}

Figure \ref{fig_multipoles} shows the scattering cross section of the core shell cluster with bulk and hollow nanospheres and the contributions of the different multipole moments. It can be recognized that the magnetic dipole resonance of bulk nanospheres is partially superimposed by higher-order resonances where the electric quadrupole contribution is of comparable strength. By contrast for thin silver nanoshells the sharper magnetic dipole resonance is well separated from the quadrupole contributions and the magnetic quadrupole resonance is additionally strongly suppressed. This allows for a stronger magnetic dipole response and a highly dispersive effective permeability of the structure.

Up to this point, we have shown that the magnetic response of core-shell clusters can be shifted to the IR regime in replacing solid nanospheres with nanoshells. The advantage is the reduced relative size of the cluster when compared to the resonance wavelength. In the best case this ratio is about $1/7$ which can be considered as sufficiently sub-wavelength. To motivate the practical realization of these clusters, we discuss in the following the effect of disorder on their response. If the performance is sufficiently tolerant against disorder, their fabrication by self-assembly techniques will be certainly attractive.

\begin{figure*}[htbp]
\centering \includegraphics[width=\textwidth]{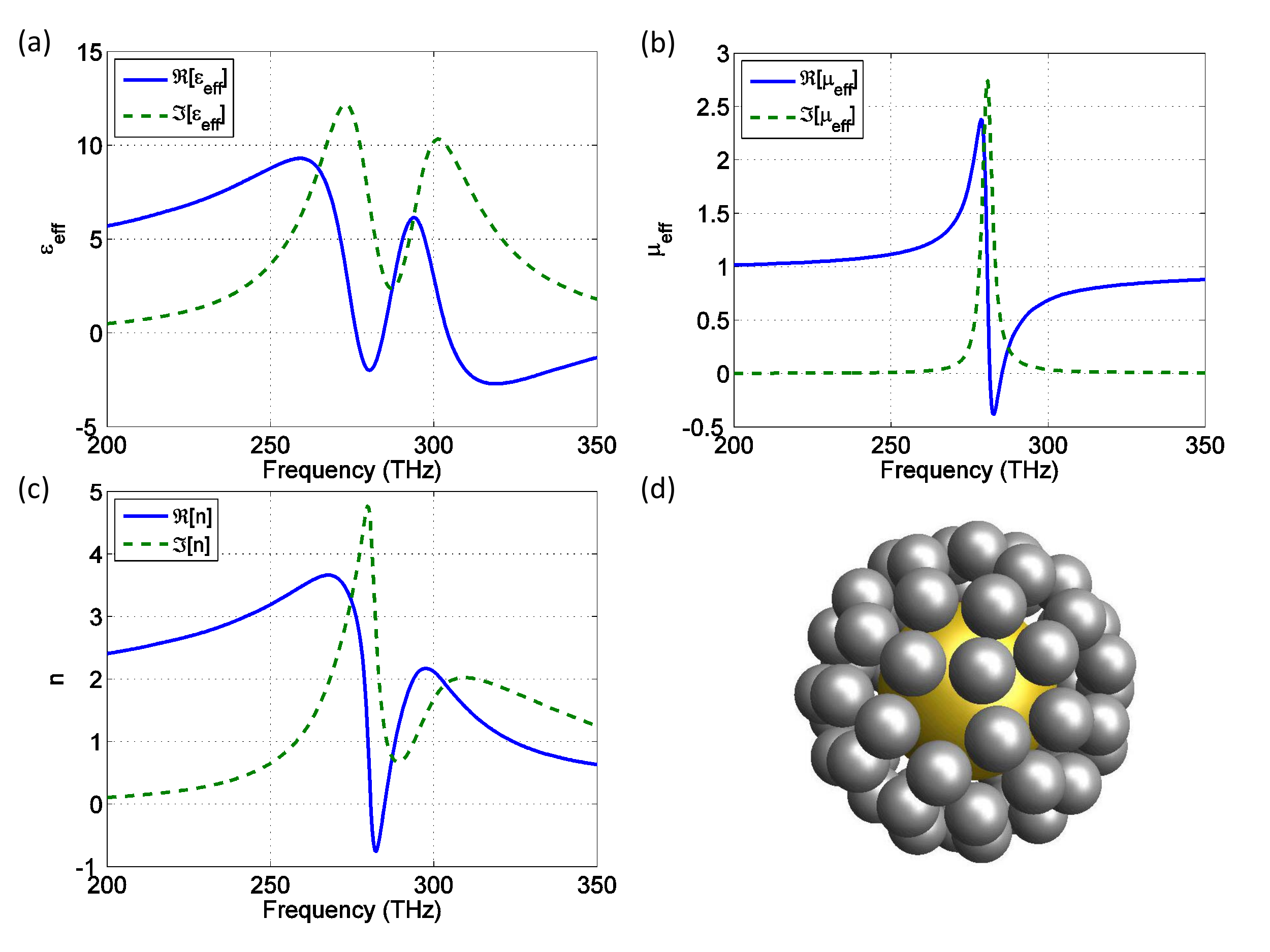} \caption{Effective permittivity (a), permeability (b)
and index of refraction (c) of a core-shell structure with $r_{\mathrm{i}}=13.8$ nm and $r_{\mathrm{o}}=15$ nm of the silver nanoshells and $r_{\mathrm{i}}=45$ nm and $r_{\mathrm{o}}=52$ nm of the central gold shell. d) Sketch of the core-shell structure.}

\label{fig_nim}
\end{figure*}

Generally, the enforced order of the nanoshells promotes the excitation of a magnetic dipole moment. This is verified by investigating the magnetic response for disordered nanoshells. Figure \ref{fig_waterfall}b) shows how disorder affects the resonance. The disorder is enforced while imposing a normally distributed spatial displacement (quantified by $\sigma$) of each nanoshell with respect to its geometrical position on the dodecahedron. The disorder causes the resonance to get slightly weaker and broader, similar to a superposition of many resonances at different frequencies. We conclude from this analysis that if the accuracy is kept approximately below $1$ nm in the fabrication process, the response degrades only marginal and a negative permeability is preserved. This requirement may be achieved with self-assembly techniques exploiting electrostatic interaction.

\section{Negative index material}

As shown in the preceding sections, a negative permeability can be achieved with the present nanoshell configuration. If the negative permeability $\M$ is accompanied by a highly dispersive permittivity a negative refractive index can occur at that frequency. This can be achieved by an extension to the previous design: instead of a dielectric sphere a single gold shell particle serves as the core. This core exhibits a strong electric dipole resonance, thus a dispersive and possibly negative permittivity. With the core-shell cluster made of nanoshell particles and the central gold shell as core, numerous parameters can be varied and both resonances can be almost independently tailored such that they almost coincide. In Fig.\ \ref{fig_nim} the effective permittivity, permeability and refractive index, calculated with $n=\sqrt{\E\M}$, of the core shell cluster are displayed. Indeed, this optimized structure shows a negative refractive index in a certain frequency domain of about $10$ THz width, with values down to about $n=-0.8$. By changing the radii of the structure, we can shift the region of negative refractive index to different frequencies to a certain extent.

\section{Conclusion}

A deep-subwavelength meta-atom that exhibits a strong magnetic dipole response has been designed. By starting from a core-shell cluster with a dodecahedron order for the solid shell nanospheres, this design has been further developed by considering nanoshells. It has been shown that the magnetic resonance can be tuned over a wide spectral domain while maintaining the size of the meta-atom. A further modification of the meta-atom where the dielectric core has been replaced by a gold shell provides additionally a strong electric dipole response. Thus, the combination of the resulting negative permeability and permittivity leads to a material with negative refractive index. Furthermore, the effect of disorder on the design has been investigated. Therewith, accurate design parameters for a possible fabrication of the presented core-shell clusters by self-assembly techniques could be derived. It is very likely that the use of metallic shells instead of solid particles constitutes a viable route to overcome some limitations of current self-assembled metamaterials. We are convinced that not just the structure presented in our work will benefit, but many other meta-atoms fabricated by self-assembly processes will equally turn out to be better in their performance while considering shells instead of solid nanospheres.

\section*{Acknowledgement}

This work was supported by the German Federal Ministry of Education and Research(PhoNa) and by the Thuringian State Government(MeMa).

\end{document}